\documentclass[letterpaper,twocolumn]{jpsj3}
\usepackage{txfonts}

\usepackage{xr}
\usepackage{graphicx}
\usepackage{amsmath,amssymb}
\usepackage{bm}
\usepackage{color}

\newcommand{\eq}[1]{(\ref{#1})}
\newcommand{\Eq}[1]{Eq.~(\ref{#1})}
\newcommand{\Figure}[1]{Figure~\ref{#1}}
\newcommand{\Fig}[1]{Fig.~\ref{#1}}
\newcommand{\Sec}[1]{Sec.~\ref{#1}}
\newcommand{\Ref}[1]{Ref.~\cite{#1}}
\def\beq{\begin{equation}} \def\eeq{\end{equation}}
\def\bea{\begin{eqnarray}} \def\eea{\end{eqnarray}}
\def\bse{\begin{subequations}} \def\ese{\end{subequations}}

   \def\vecr{{\bm r}} 
    
\def\||{\parallel}

\def\<{\left\langle} \def\>{\right\rangle}
\def\({\left(} \def\){\right)}

\title{Is a Doubly Quantized Vortex Dynamically Unstable in Uniform Superfluids?}

\author{Hiromitsu Takeuchi$^1$\thanks{hirotake@sci.osaka-cu.ac.jp}, Michikazu Kobayashi$^2$, and Kenichi Kasamatsu$^3$}
\inst{$^1$Department of Physics, Osaka City University, 3-3-138 Sugimoto, Sumiyoshi-ku, Osaka 558-8585, Japan \\
$^2$Department of Physics, Kyoto University, Oiwake-cho, Kitashirakawa, Sakyo-ku, Kyoto 606-8502, Japan \\
$^3$Department of Physics, Kindai University, Higashi-Osaka, Osaka 577-8502, Japan} 

\abst{
We revisit the fundamental problem of the splitting instability of a doubly quantized vortex in uniform single-component superfluids at zero temperature.
We analyze the system-size dependence of the excitation frequency of a doubly quantized vortex through large-scale simulations of the Bogoliubov--de Gennes equation,
and find that the system remains dynamically unstable even in the infinite-system-size limit.
 Perturbation and semi-classical theories reveal that the splitting instability radiates a damped oscillatory phonon as an opposite counterpart of a quasi-normal mode.
}

\begin{document}
\maketitle

\textit{Introduction}: 
Vortices appear in many branches of physics. In particular, the structure, stability, and dynamics of 
vortices in nonlinear fields share common features in many physical systems \cite{Pismen}. 
Quantized vortices are prototypes among those vortices, playing a key role in the fluid dynamics of superfluid helium and Bose--Einstein condensates (BECs) \cite{Donnellybook, 2008Pethick, 2013Tsubota, Pitaevskiibook}. 
In general, quantized vortices are characterized by the winding number of the phase of the superfluid order parameter around the vortex core. A vortex whose winding number $l$ is more than unity is called an $l$-quantized or multiply quantized vortex (MQV).
Since the energy of an $l$-quantized vortex is generally larger than the sum of energies of $l$ singly quantized vortices (SQVs), an MQV is energetically unstable and splits into SQVs in uniform systems \cite{VortexPotential}.
In fact, MQVs have never been observed in equilibrium.
However, this argument does not eliminate the possibility that MQVs survive as a metastable state at very low temperatures
 when energy dissipation is negligible.

To investigate the splitting instability precisely, we need 
to analyze the microscopic structure of the vortex core. 
It is difficult to demonstrate such an analysis in the strongly correlated superfluid $^4$He.
Experimentally, there is no established technique to prepare an MQV 
in helium superfluids as an initial state of the instability problem.
The realization of MQV in the BECs of ultra-cold gases sheds light on this problem, and vortex splitting has been observed \cite{2004Shin,2007Isoshima,2010Kuwamoto}.
The MQV in trapped systems can be dynamically unstable, and split into vortices with smaller winding numbers according to the Bogoliubov--de Gennes (BdG) analysis at zero temperature \cite{1999Pu,2000Skryabin,2003Mottonen,2004Kawaguchi,2006Huhtamaki,2006Lundh,2007Fukuyama,2008Nilsen}.
Dynamic instability may occur when the excitation modes have complex frequencies as a result of coupling or ``mixing'' between two modes with positive and negative excitation energies.
 The negative energy mode, called the core mode, is localized at the vortex core and decreases the angular momentum of the system by $-l\hbar$ in the direction along the core.
The positive energy mode is a collective mode of the condensate. 
The instability depends on the atomic interaction strength in a complicated manner \cite{1999Pu,2000Skryabin,2003Mottonen,2004Kawaguchi,2006Huhtamaki,2006Lundh},
obfuscating the underlying physics.
Lundh and Nilsen made progress in understanding the splitting instability by
employing a perturbation theory;
 however, no quantitative evaluation was carried out because of the complicated behavior of the imaginary part of excitation frequency (see Fig.~3 in \Ref{2006Lundh}).

Currently, we do not have a definite answer to the question ``does the splitting instability occur in \textit{bulk superfluids at zero temperature}?".
This is partly because long-time numerical simulations with high-spec computers are required for investigating the dynamic stability more precisely.
According to the previous studies on trapped BECs 
\cite{1999Pu,2000Skryabin,2003Mottonen,2004Kawaguchi,2006Huhtamaki,2006Lundh,2007Fukuyama,2008Nilsen}, it is not easy to answer this question,
 because the finite-size-effect is essential there. 
Although Aranson and Steinberg \cite{1996Aranson} concluded that
 the lifetime of an MQV may become infinite without a trap in their numerical simulation, its system-size dependence has not been clarified systematically. This problem is of fundamental importance in quantum fluid dynamics at very low temperatures, and therefore, it is essential to understand, {\it e.g.,} quantum turbulence of helium superfluid \cite{2002vinen} and large two-dimensional (2D) BECs \cite{2014White},
 where the problems become more complicated if the presence of MQVs is permitted.
 
Here, we consider the most fundamental situation of a doubly quantized vortex (DQV) in a uniform 2D system.
We show that a DQV is dynamically unstable in uniform BECs. 
Our large-scale numerical computation of the BdG equations reveals a nontrivial system-size dependence of the excitation frequency and its asymptotic behavior in the infinite-system-size limit.
The nontrivial dependence is well-characterized by the ``mixing'' between the core mode and phonon with our rescaling perturbation theory.
The semi-classical theory, extended to the case of complex eigenvalue, reveals that the instability causes spontaneous radiation and amplification of quasi-normal modes such as the damped oscillatory phonons with anomalously long attenuation length.
We discuss an analogy between this phenomenon and the rotational superradiance, which was observed recently as an amplification of surface water waves by a draining vortex \cite{2017Torres}.

{\it Formulation}: 
We consider BECs in a quasi-2D system at zero temperature 
when the degrees of freedom along the $z$-axis are not considered \cite{Zaxis}.
The condensate is well-described 
by the order parameter $\psi(\vecr,t)$, which obeys the Gross--Piteavskii (GP) Lagrangian
$
{\cal L}=\int d^2x \psi^*\(  i\hbar\partial_t - H - \frac{g}{2}|\psi|^2 \)\psi
$.
Here, we use $H=-\hbar^2 \nabla^2/(2m_{\rm a}) -\mu$ with the atomic mass $m_{\rm a}$, 
the chemical potential $\mu$, and the interaction constant $g$.

Without loss of generality, a DQV with positive winding number $l=2$ is considered.
The stationary state of a DQV is written as $\psi(\bm{r},t) = \phi(\vecr)=f(r) e^{il\theta}$ with the cylindrical coordinates $\vecr=(r,\theta)$.
The real amplitude $f(r)$ obeys the GP equation 
$[ H_r+l^2\hbar^2 /(2m_{\rm a} r^2)+g f^2 ] f = 0$
with $H_r=-\hbar^2 \( \partial_r^2 + r^{-1} \partial_r  \) /(2m_{\rm a}) - \mu$.
The dimensionless amplitude $\tilde{f}=\sqrt{g/\mu}f$ is characterized by the rescaled length $\tilde{r}=r/\xi$ with the healing length $\xi=\hbar/\sqrt{m_{\rm a}\mu}$;
$\tilde{f}$ approaches the asymptotic form $\tilde{f}^2 \approx  1 -  l^2/(2\tilde{r}^2) $ for $\tilde{r} \gtrsim 1$ and $\tilde{f}^2 \propto \tilde{r}^{2l}$ for $\tilde{r} \to 0$.
In other words,
our system is parameterized only by the effective system size $\tilde{R}=R/\xi$ through the boundary condition at $r=R$.
The same statement is applicable to the BdG analysis below.

To investigate the stability, we introduce a fluctuation 
$\delta\psi (\vecr,t)=\psi (\vecr,t)-\phi (\vecr)=e^{2i\theta}\left[ u(r)e^{im\theta-i\omega t}-v^*(r)e^{-im\theta+i\omega^* t} \right]$. 
The linearization of the equation of motion of the Lagrangian ${\cal L}$ with respect to $\delta\psi$ leads to the BdG equation for
 $\vec{u}=(u,v)^T$,
\begin{eqnarray}\label{Eq_radialBdG}
\hbar\omega \vec{u}=\hat{h} \vec{u} \equiv \left[
\begin{array}{cc}
h_+ & -gf^2 \\
gf^2 & -h_- \\
\end{array} 
\right] \vec{u}
\end{eqnarray}
with $h_{\pm}=H_r+\frac{\hbar^2(l\pm m)^2}{2m_{\rm a}r^2}+2gf^2$.
The excitation energy $\epsilon_\alpha$ for an eigensolution $\(\omega, \vec{u} \)=\(\omega_\alpha, \vec{u}_\alpha \)$ is written as $\epsilon_\alpha=\hbar \omega_\alpha {\cal N}_{\alpha\alpha}$.
Here, the excitations are labeled by the integer $\alpha$, and the norm ${\cal N}_{\alpha\beta}=2\pi \int_0^\infty dr r \vec{u}_\alpha^\dagger \hat{\sigma}_z\vec{u}_\beta =\pm \delta_{\alpha\beta}$ is defined for real eigenvalues with a matrix $\hat{\sigma}_z={\rm diag}(1,-1)$ and the Kronecker's delta $\delta_{\alpha\beta}$.
From the orthogonality relation $( \omega_\alpha-\omega_\beta^* ) {\cal N}_{\alpha\beta}=0$ \cite{2008Pethick},
$\epsilon_{\alpha}$ becomes zero with ${\cal N}_{\alpha\alpha}=0$ for ${\rm Im}(\omega_\alpha)\neq 0$.
The vortex state is dynamically unstable when there is at least one eigensolution with ${\rm Im}(\omega)>0$, 
because the corresponding excitation is amplified exponentially as $\propto e^{{\rm Im}(\omega)t}$.

\begin{figure}
\begin{center}
\includegraphics[width=1.0 \linewidth,keepaspectratio]{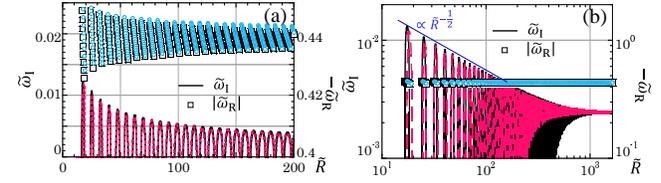}
\end{center}
\caption{(Color online) 
Dependence of the dimensionless eigenvalue of the splitting instability on $\tilde{R}$ for $\tilde{R}\leq 200$ (a) and the double logarithmic plot for $\tilde{R}\leq 1638.4$ (b).
Solid  (dashed) curves and squares (circles) represent the numerical (analytical) result of $\tilde{\omega}_{\rm I}$ and $\tilde{\omega}_{\rm R}$, respectively.
The analytical result is obtained by \Eq{omega_per} with formula \eq{Eq_Wmix_int}, presumed from the overall $\tilde{R}$-dependence of $\tilde{\omega}_{\rm I}$.
} 
\label{Fig_R_w}
\end{figure}

{\it Numerical results}:
The eigenvalue problem is numerically analyzed for a cylindrical system of 
dimensionless radius $\tilde{R}$ by diagonalizing \Eq{Eq_radialBdG} with the numerical solution of $\tilde{f}$ under the Neumann boundary condition at $\tilde{r}=\tilde{R}$.
\Figure{Fig_R_w} shows the $\tilde{R}$-dependence of 
the dimensionless frequency
$\tilde{\omega}\equiv \hbar \omega/\mu$
 of the instability mode with $m=-l=-2$ when the imaginary part
$\tilde{\omega}_{\rm I} \equiv {\rm Im}(\tilde{\omega}) >0$ takes the largest value \cite{Ntech}.
The real part is always negative; $\tilde{\omega}_{\rm R} \equiv {\rm Re}(\tilde{\omega})<0$ for $\tilde{\omega}_{\rm I}>0$.

The eigenvalue is strongly sensitive to $\tilde{R}$, showing a nearly periodic behavior below $\tilde{R} \sim 500$.
A similar behavior can be seen in the numerical results for trapped BECs \cite{1999Pu,2000Skryabin,2003Mottonen,2004Kawaguchi,2006Huhtamaki,2006Lundh}.
In these studies, the excitation frequency was parameterized by the interaction strength or particle number, which made the analysis of the problem complicated.

Figure~\ref{Fig_R_w}(b) shows that the peak values of $\tilde{\omega}_{\rm I}$ are 
proportional to $\tilde{R}^{-1/2}$ for $\tilde{R} \lesssim 500$,
while $\tilde{\omega}_{\rm I}$ is asymptotic to a finite value for $\tilde{R} \gtrsim 500$. 
This fact indicates that a DQV is dynamically unstable 
even in the infinite-system-size limit $\tilde{R}\to \infty$.
We estimated the values of $\tilde{\omega}_{\rm R}$ and $\tilde{\omega}_{\rm I}$ in this limit as 
$\tilde{\omega}_{\rm R} \to \tilde{\Omega}_\infty= -0.438969(2)$ and
 $\tilde{\omega}_{\rm I} \to 1/\tilde{\tau}_\infty=  0.002429(2)$
 \cite{Extrapolating}.
Here,  $\tilde{\Omega}_\infty$ corresponds to the angular frequency $\Omega_\infty=\tilde{\Omega}_\infty\mu/\hbar$ of the rotational motion of the two parallel SQVs into which a DQV splits,
 and $\tilde{\tau}_\infty$ characterizes the growth time $\tau_\infty=\tilde{\tau}_\infty \hbar/\mu$ of the distance between the SQVs.

From the naive consideration given below,
we can expect that the splitting instability should be less sensitive to $\tilde{R}$ for $\tilde{R}\to \infty$.
According to the theory of Hamiltonian dynamical systems \cite{1987MacKay}, 
the dynamic instability can be induced by a mixing between positive and negative energy modes. 
Here, the negative energy mode corresponds to the core mode localized around the vortex core ($\tilde{r} \lesssim 1$).
In trapped systems, the positive energy mode may correspond to a collective mode that causes a ripple wave along the surface of the condensate.
However, such a surface-localized mode cannot correlate with the core mode in the infinite-system-size limit.
On the other hand,
phonons can play the role of the positive energy mode,
because its wave function is distributed broadly within the bulk and its correlation to the core mode may remain even for $\tilde{R} \to \infty$ in our problem.

{\it Rescaling perturbation analysis}:
To describe our problem quantitatively,
we introduce a perturbation analysis for the BdG equations in a different manner from that in \Ref{2006Lundh}.
We parameterize the chemical potential and the interaction coefficient with a perturbation parameter $\lambda(\ll 1)$ as $\mu=\mu_0 (1+\lambda )$ and $g \to g_\lambda\equiv g_0(1+\lambda)$, respectively.
Then, the effective system size is represented as
$\tilde{R}=\tilde{R}_0 \sqrt{1+\lambda }$
through $\xi=\xi_0/\sqrt{1+\lambda}$.
To express the $\lambda$-dependence explicitly,
 we write $f\to f_\lambda$ and $\hat{h}\to\hat{h}_\lambda$ in the following.

The perturbation for the BdG equations is represented by the deviation
\begin{eqnarray}\label{Eq_BdG_per}
\delta\hat{h}=\hat{h}_\lambda-\hat{h}_0= \lambda \left[
\begin{array}{cc}
-\mu_0+2 G' &  - G' \\
 G' & + \mu_0 - 2 G' \\
\end{array} 
\right]
\end{eqnarray}
with
$G'=\lim_{\lambda \to 0} \frac{g_\lambda f_\lambda^2 -g_0 f_0^2}{\lambda}
 \to \partial_\lambda \( g_\lambda f_\lambda^2 \)$.
Suppose that the system becomes dynamically unstable when $\lambda$ increases from the unperturbed state ($\lambda=0$) to a perturbed state ($\lambda \neq 0$).
The eigenvector $\vec{u}$ in the perturbed state is described by a linear combination of the eigenvectors $\vec{u}_\alpha$ ($\alpha=1,2,...$) in the unperturbed state as 
$
\vec{u}=\sum_\alpha {\cal C}_\alpha \vec{u}_\alpha
$.
The coefficient vector ${\bm C}=\left[ {\cal C}_1, {\cal C}_2,\cdots \right]^T$ obeys the eigenvalue equation 
$\hbar \omega {\bm C}=\( \check{\cal H}_0+\lambda \check{\cal W} \) {\bm C}$ 
with  $\check{\cal H}_0= {\rm diag}. \( \hbar\omega_1,\hbar\omega_2, \cdots \) $ and
$
\left[ \check{\cal W} \right]_{\alpha\beta}={\cal W}_{\alpha\beta}\equiv \frac{2\pi}{\lambda {\cal N}_{\alpha\alpha}}
 \int_0^\infty dr r \vec{u}_\alpha^\dagger \hat{\sigma}_z \delta\hat{h} \vec{u}_\beta.
$

When the frequencies of the two modes, namely, $\alpha=1,2$, are very close to each other, we may apply the two-mode approximation as in conventional quantum mechanics, {\it i.e.}, only the contribution from the two modes is considered while neglecting that from all the other modes \cite{PtheoryLandau}.
 Dynamic instability can occur in the case of ${\cal N}_{11}{\cal N}_{22}=-1$ \cite{2000Skryabin, 2006Lundh}.
 Consider that the phonon and core mode correspond to $\alpha=1$ and $2$, respectively.
 The core mode, whose angular momentum is $-l\hbar$, should have the positive norm ${\cal N}_{22}=1$ for our case of $l=-m=2$,
 since the angular momentum carried by the $\alpha$ mode is given by $m \hbar{\cal N}_{\alpha\alpha}$.
Then, the eigenvalue for ${\bm C}=\left[ {\cal C}_1, {\cal C}_2 \right]^T$ reduces to
\begin{eqnarray}\label{omega_per}
\tilde{\omega}=\(\tilde{\varepsilon}_{\rm core}-\tilde{\varepsilon}_{\rm ph}\)/2 
\pm i\sqrt{\tilde{W}_{\rm mix}^2-\(\tilde{\varepsilon}_{\rm ph}+\tilde{\varepsilon}_{\rm core}\)^2/4 }. 
\end{eqnarray}
Here, we used $\tilde{\varepsilon}_{\rm ph}=\tilde{\varepsilon}_1$ and $\tilde{\varepsilon}_{\rm core} = \tilde{\varepsilon}_2$
 with the dimensionless form $\tilde{\varepsilon}_\alpha={\cal N}_{\alpha\alpha}\( \hbar \omega_\alpha+\lambda {\cal W}_{\alpha\alpha} \) /\mu$ of the perturbed excitation energy without taking the mixing interaction $\tilde{W}_{\rm mix}^2 \equiv  \lambda^2 \left| {\cal W}_{12}/\mu\right|^2 $ into account.

The phonon dispersion is represented by a function of $\tilde{R}$ as
$\tilde{\varepsilon}_{\rm ph}(\tilde{k}_j)=\tilde{k}_j ( 1+\tilde{k}_j^2/4  )^{1/2}$ with the quantized wave number $\tilde{k}_j=\pi (j+1/2) /\tilde{R}$, where $j$ is the number of nodes in the radial direction.
Here, the adjustment $1/2$ was introduced by considering the boundary condition $v(\tilde{r}=0)=0$ for $m=-2$.
The number $j$ is chosen such that $\tilde{\varepsilon}_{\rm ph}(\tilde{k}_j)$ gives the closest value to $-\tilde{\varepsilon}_{\rm core}$ for a given $\tilde{R}$.
The dimensionless energy $\tilde{\varepsilon}_{\rm core}$ of the core mode
 is independent of $\tilde{R}$ for $\tilde{R} \gg 1$ according to the naive consideration above.
In fact, $\tilde{\varepsilon}_{\rm core}$ has been computed numerically by an approximate method in \Ref{2006Lundh} as $\tilde{\varepsilon}_{\rm core}=-0.439$, to which $\tilde{\Omega}_\infty$ reduces, as we shall explain below.

\begin{figure}
\begin{center}
\includegraphics[width=1.0 \linewidth,keepaspectratio]{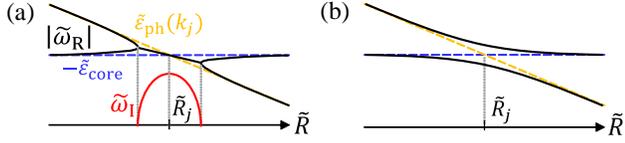}
\end{center}
\caption{(Color online) 
Schematics of a bubble for the splitting instability (a) and the so-called avoided crossing (b).
} 
\label{Fig_Mix}
\end{figure}

\Figure{Fig_Mix}~(a) shows the schematic of the $\tilde{R}$-dependence of $\tilde{\omega}$ described by \Eq{omega_per}.
The imaginary part appears around $\tilde{R}=\tilde{R}_j \equiv \pi (j+1/2)/\tilde{k}_{\rm core}$,
 at which $\tilde{\omega}_{\rm I}$ takes a local maximum value $|\tilde{W}_{\rm mix}|$ with $\tilde{\varepsilon}_{\rm ph}(\tilde{k}_{\rm core})=-\tilde{\varepsilon}_{\rm core}$.
This structure is called a {\it bubble} of instability \cite{1987MacKay}.
The real part $\tilde{\omega}_{\rm R}=\(\tilde{\varepsilon}_{\rm core}-\tilde{\varepsilon}_{\rm ph}\)/2$ is negative in a bubble with $\tilde{\varepsilon}_{\rm core}<0$ and $\tilde{\varepsilon}_{\rm ph}>0$.

 In the numerical plots shown in \Fig{Fig_R_w},
 we can see that many bubbles similar to the structure of \Fig{Fig_Mix}~(a) appear periodically around $\tilde{R}=\tilde{R}_j$.
The neighboring bubbles overlap for large $\tilde{R}$,
 and the width of a region with $\tilde{\omega}_{\rm I} = 0$ becomes smaller and disappears as $\tilde{R}$ increases \cite{BubbleWidth}.
 Finally, the asymptotic values of $\tilde{\omega}_{\rm R,I}$ are given by \Eq{omega_per} with $\tilde{\varepsilon}_{\rm ph}(\tilde{k}_{\rm core})=-\tilde{\varepsilon}_{\rm core}$.
 Therefore, we may write
 $\tilde{\Omega}_\infty \to \tilde{\varepsilon}_{\rm core}$ and $1/\tilde{\tau}_\infty \to |\tilde{W}_{\rm mix}|$.

We can deduce the power-law behavior of $\tilde{W}_{\rm mix}$ from the overall profile of $\tilde{\omega}_{\rm I}$ in \Fig{Fig_R_w}~(b);
 $\tilde{W}_{\rm mix}\propto \tilde{R}^{1/2}$ for $\tilde{R} \lesssim 500$ (solid line) and $\tilde{W}_{\rm mix} \approx 1/\tilde{\tau}_\infty$ for $\tilde{R} \gtrsim 500$.
 Such a behavior is anomalous in the sense
 that the length $R\sim 500 \xi$, around which the power-law behavior changes, is irrelevant to any possible scale in our formalism described above \cite{PowerLaw}.

{\it Extended semi-classical analysis}:
To demonstrate the anomalous behavior beyond the perturbation analysis,
we introduce the semi-classical theory for the BdG equations,
 which can be used to describe low-energy modes far from a topological defect or an interface \cite{2013Takeuchi,2015Takahashi}. 
Here, we extend the theory to our case with complex excitation frequencies.

The semi-classical theory starts from the Wentzel--Kramers--Brillouin (WKB) ansatz for the excitation wave function $\vec{u}$ in the first-order approximation,
$\vec{u}(r) =e^{\frac{i}{\hbar}S} \vec{U}$
 with $S(r)=S_0(r)+\frac{\hbar}{i}S_1(r)$ and $\vec{U} = (U,V)$. 
Substituting the ansatz into \Eq{Eq_radialBdG},
 we obtain
\begin{eqnarray}\label{Eq_BdG_WKB}
E \vec{U}=\left[
\begin{array}{cc}
E_+ & -gf^2 \\
gf^2 & -E_- \\
\end{array} 
\right] \vec{U}+\frac{\hbar}{i} D\hat{\sigma}_z  \vec{U},
\end{eqnarray}
where we use $E_\pm=\frac{P_r^2}{2m_{\rm a}}+\frac{\(M\pm L\)^2}{2m_{\rm a}r^2}
+2gf^2-\mu$, $(E, M, L)= (\hbar\omega,\hbar m,\hbar l)$,  $P_r=\frac{dS_0}{dr}$,
and $D(r)=\frac{P_r}{m_{\rm a}}\frac{dS_1}{dr}+\frac{1}{2m_{\rm a}}\(\frac{dP_r}{dr}+\frac{P_r}{r}\)$.
The zeroth-order approximation, called the classical limit, neglects the second term on the right hand side of \Eq{Eq_BdG_WKB}, 
yielding $\( E - E_+ \) \( E + E_- \)+g^2f^4=0$.
The first-order correction reduces to the relation $dS_1/dr= -(2r)^{-1}$.

Considering the bulk region, which is far from the vortex core ($r \gg \xi$),
 and neglecting the terms of $O\( \xi^2/r^2 \)$,
we have $E^2 = \frac{P_r^2}{2m_\mathrm{a}} \left(  \frac{P_r^2}{2m_\mathrm{a}}  + 2 \mu \right) $ \cite{SCtheory}.
For our case of a complex eigenvalue $E/\mu=\tilde{\omega}_{\rm R}+i\tilde{\omega}_{\rm I}$ with $|\tilde{\omega}_{\rm R}| \gg \tilde{\omega}_{\rm I} \geq 0$,
the radial momentum $P_r$ is written as $P_r\xi/\hbar=\tilde{k}+i \tilde{\kappa}$ with $\tilde{\kappa} \ll \tilde{k}$.
Subsequently, we obtain
\begin{eqnarray}\label{Eq_k_kappa}
\tilde{k} = - |\tilde{\omega}_{\rm R}|\( 1- \tilde{\omega}_{\rm R}^2/8 \),~~
\tilde{\kappa} = \tilde{\omega}_{\rm I} \(1-3\tilde{\omega}_{\rm R}^2/8 \)
\end{eqnarray}
 up to the order of $O\( \tilde{\omega}_{\rm R}^4 \)$. 
 Here, the sign of $\tilde{k}$ is negative since the outgoing phonon has ${\cal N}_{11}<0$ in our perturbation analysis.
As a result, we obtain
$
S(r)  = \hbar \tilde{k} \tilde{r}+i\hbar \tilde{\kappa}\tilde{r}-\frac{\hbar}{2i}\ln \tilde{r}+ {\rm const}.
$


\begin{figure}
\begin{center}
\includegraphics[width=1.0 \linewidth,keepaspectratio]{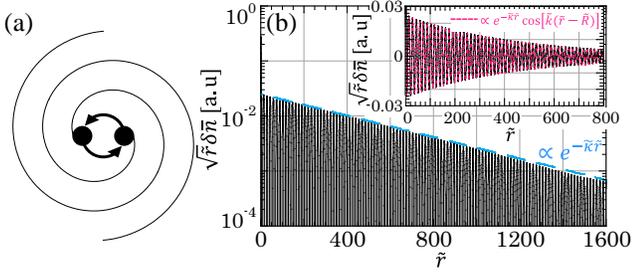}
\end{center}
\caption{(Color online) (a) Schematic of the wavefront of radiated phonon in the splitting instability of a DQV. Black circles represent two parallel SQVs. 
(b) The cross-section profile $\sqrt{\tilde{r}}\delta \bar{n}$ of the density fluctuation induced by the instability mode for $\tilde{R}=1638.4$.
 The dashed line represents the overall damping $\propto e^{-\tilde{\kappa}\tilde{r}}$.
The inset shows the magnified image of the numerical plot (solid curve) and the analytical plot of \Eq{Eq_dn} (dashed curve). 
The values of $\tilde{k}$ and $\tilde{\kappa}$ are given by \Eq{Eq_k_kappa} with the numerical data of $\tilde{\omega}_{\rm R}$ and $\tilde{\omega}_{\rm I}$ in \Fig{Fig_R_w}. 
} 
\label{Fig_r_u}
\end{figure}

To demonstrate the accuracy of our theory,
we describe an observable quantity, {\it i.e.}, the density fluctuation $\delta n (r,\theta,t) \equiv \left| \psi (\bm{r},t) \right|^2-\left| \phi(\bm{r}) \right|^2$ induced by the instability mode.
The semi-classical solution gives
 $\delta n \approx 2 {\rm Re}(\phi^*\delta\psi)
=   \frac{2f\left| U-V \right|}{\sqrt{\tilde{r}}}e^{-\tilde{\kappa} {r}+\omega_{\rm I}t}  \cos ( m\theta - \omega_{\rm R}t-\tilde{k}\tilde{r} + \Theta )$ with a constant $\Theta$ [also see Fig.~\ref{Fig_r_u}(a)].
For simplicity, we evaluate the semi-classical result for the cross-section profile
\begin{eqnarray}\label{Eq_dn}
\delta \bar{n}(r)\equiv \delta n (r,0,0)  \propto \tilde{r}^{-1/2}  e^{-\tilde{\kappa}\tilde{r}} \cos \left[ \tilde{k} \(\tilde{r}-\tilde{R}\) \right].
\end{eqnarray}
Here, we took the boundary conditions at $\tilde{r}=\tilde{R}$ into account.
In Fig.~\ref{Fig_r_u}(b), we compare the radial profile $\sqrt{\tilde{r}}\delta\bar{n}(r)$ of the numerical solution with Eq.~(\ref{Eq_dn}),
obtaining a good agreement between them for $\tilde{r}\gg 1$.

The semi-classical analysis suggests that the overall $\tilde{R}$-dependence of $\tilde{\omega}_{\rm I}$ is characterized by the rescaled damping rate in the infinite-system-size limit,
$\tilde{\kappa}_\infty \equiv \tilde{\kappa}(\tilde{\omega}_{\rm R}=\tilde{\Omega}_\infty,\tilde{\omega}_{\rm I}=\tilde{\tau}_\infty^{-1})$.
If the rescaled attenuation length $\tilde{\kappa}_\infty^{-1}$ is much smaller than $\tilde{R}$,
the boundary effect is negligible so that the instability is independent of $\tilde{R}$.
This consideration is helpful in constructing an analytic formula for the mixing interaction in order to describe the overall $\tilde{R}$-dependence.
We found the simplest interpolating formula between the two limits $\tilde{\kappa}_\infty \tilde{R} \to 0$ and $\tilde{\kappa}_\infty \tilde{R} \to \infty $, 
\begin{eqnarray}\label{Eq_Wmix_int}
\tilde{W}_{\rm mix}^2= \tilde{\tau}_\infty^{-2} \left[1+1/( \tilde{\kappa}_\infty \tilde{R} )^2 \right]^{1/2}. 
\end{eqnarray}
The complex frequency of Eq.~\eqref{omega_per} with \Eq{Eq_Wmix_int} describes the numerical result very well, as shown in \Fig{Fig_R_w}.

Finally, we make a physical interpretation of such an anomalous damped oscillatory mode by regarding it as an opposite counterpart of a quasi-normal mode,
 which is typically discussed in the context of gravitational waves from a perturbed black hole (BH) \cite{1999Kokkotas}.
A perturbed BH evolves into the unperturbed spherical shape by decreasing its asymmetry exponentially in time;
the deviation from a spherical shape is proportional to $e^{- t/\tau_{\rm B}}$ with the decay time $\tau_{\rm B} >0$.
In this process, the radiated gravity wave is described as the formal solution of a Schr\"odinger-like equation, whose eigenvalue becomes complex through the boundary effect of the BH, namely, the quasi-normal mode.
 In the WKB approximation, the wave forms a growing oscillation with the growth rate $\kappa_{\rm B}=(c\tau_{\rm B})^{-1}$,
where $c$ is the speed of light far from the BH.
On the contrary, in the case of splitting instability,
a DQV is perturbed to split into two SQVs by increasing the asymmetry of the ``oscillatory source'' exponentially in time;
$d \propto e^{ t/\tau_\infty}$.
The radiated phonon produces a damped oscillation with the damping rate $\kappa_\infty \approx (c_{\rm s} \tau_\infty)^{-1}$ characterized by the speed $c_{\rm s}=\sqrt{\mu/m}$ of phonon,
 which is consistent with \Eq{Eq_k_kappa} in the linear dispersion approximation, $|\tilde{\omega}_{\rm R}|\approx \tilde{k}$.
The relation between dynamic instability and quasi-normal modes has been also discussed in the context of BH physics in \Ref{2016Coutant}.

{\it Discussion}:
The radiation of the quasi-normal mode in the splitting instability produces a double spiral density wave [\Fig{Fig_r_u}(a)] in the early stage of instability development.
We have no satisfactory explanation of why $1/\tilde{\tau}_\infty$ is so small,
although it might be related to the vortex-vortex interaction potential.
The quasi-normal mode can be observed experimentally in highly oblate BECs whose size is much larger than the healing length.
The instability is induced by an external perturbation, {\it i.e.,} an external optical potential that violates the rotational symmetry of the initial state with a DQV.

An incident plane wave of phonon, whose energy $\mu\tilde{\varepsilon}_{\rm ph}$ is close to  $-\mu\tilde{\varepsilon}_{\rm core}$, may trigger the instability.
Then, the incident phonon will be amplified due to the exponential growth of the instability mode.
This phenomenon is analogous to the rotational superradiance that has been observed recently in classical fluids \cite{2017Torres},
 where the incident waves on the water surface are amplified by a draining vortex.
The experiment in the dissipative classical fluid system did not reveal the mechanism behind the negative energy mode,
 which exists along with a positive energy mode to obey the energy conservation law. 
In our  isolated quantum fluid system,
 the superradiant amplification is caused by the pair nucleation of positive and negative energy modes;
 the latter is represented by the core mode as a {\it bound state} whose existence is classically limited inside the so-called ``ergo region'' $r<r_{\rm e}$, and the former can propagate outside.
Here, $r_{\rm e}/\xi=\sqrt{|lm|/\tilde{\varepsilon}_{\rm ph}}$ is the effective ergo radius at which the semi-classical energy becomes zero as $E \approx \mu\tilde{\varepsilon}_{\rm ph}+\hbar m \Omega(r_{\rm e})=0$ with $\tilde{\varepsilon}_{\rm ph}=-\tilde{\varepsilon}_{\rm core}$ and the local superfluid velocity $r\Omega(r)=\frac{\hbar l}{m_{\rm a} r}$.
Hence, our system could be useful for simulating BH physics,
while a similar analogy has been discussed by considering a rotating object or a vortex in superfluid systems \cite{1999Calogeracos,2003Volovik, 2005Slatyer, 2006Federici, 2008Takeuchi}.



\begin{acknowledgments}
We are grateful to S. Inouye, H. Ishihara, M. Kimura, Y. Kawaguchi, T. Kuwamoto, and T. Mizushima for useful discussions and comments on this work.
This work was supported by KAKENHI from the Japan Society for the Promotion of Science (Grants No. 17K05549, 17H02938, 26870295, 26400371).
The present research was also supported in part by the Osaka City University (OCU) Strategic Research Grant 2017  for young researchers.
\end{acknowledgments}

\newpage
\section*{Supplemental material}

\renewcommand{\thesubsection}{A\arabic{subsection}}
\setcounter{subsection}{0}

\renewcommand{\theequation}{A\arabic{equation}}
\setcounter{equation}{0}
\renewcommand{\thefigure}{A\arabic{figure}}
\setcounter{figure}{0}

\subsection{The intervortex potential}\label{ASec:IntV}
\subsubsection*{Definition of the interaction potential}
Since we are interested in the stability of a doubly quantized vortex (DQV),
we here discuss the interaction potential between two singly quantized vortices (SQV) in uniform Bose--Einstein condensates (BECs). 
The results obtained with different methods are summarized in \Fig{fig:vortex-energy}.

First, we define the interaction energy between two vortices.
We introduce the energy ${\cal E}(d)$ of the state $\Psi_d$ with two SQVs as a function of distance $d$ between the two vortices,
\begin{eqnarray}
{\cal E}(d)=\int d^2x \Psi_d^*\(  -\frac{\hbar^2}{2m_{\rm a}} \nabla^2 + \frac{g}{2}|\Psi_d|^2 \)\Psi_d.
\nonumber
\end{eqnarray}
Measuring the energy from the sum of the energy ${\cal E}_{\rm bulk}=\frac{\mu^2}{2g}\pi R^2$ of the bulk state and that ${\cal E}_2$ of a DQV,
the interaction energy ${\cal E}_{\rm int}(d)$ between two SQVs may be defined by
\begin{eqnarray*}
{\cal E}_{\rm int}(d)\equiv {\cal E}(d)-{\cal E}_2-{\cal E}_{\rm bulk}.
\end{eqnarray*}

\subsubsection*{Estimations in the hydrodynamic and Pad\'e approximations}
Conventionally, the dimensionless energy $\tilde{\cal E}_l = {\cal E}_l g/ (\mu^2\xi^2)$ of a $l$-quantized vortex is estimated as
\begin{eqnarray}
\tilde{\cal E}_l \simeq 
\pi l^2 \log \tilde{R},
\label{eq:l-vortex-energy}
\end{eqnarray}
where we approximate the dimensionless order parameter $\tilde{\psi} = \sqrt{\mu / g} \psi$ for a $l$-quantized vortex as $\tilde{\psi}_l(0)$ with $\tilde{\psi}_l(\tilde{x},\tilde{y}) \simeq e^{i l \theta(\tilde{x})}$ and $\theta (\tilde{x})\equiv \arctan \(\frac{\tilde{y}}{ \tilde{x} }\)$.
For integration, $\tilde{r} = 1$ and $\tilde{r} = \tilde{R}$ are chosen as the lower and upper cutoffs, respectively.
Equation \eq{eq:l-vortex-energy} shows that the energy $\tilde{\cal E}_l = l^2 \tilde{\cal E}_1$ for $l$-quantized vortex is larger than the energy $l \tilde{\cal E}_1$ for $l$ SQVs.
In the similar approximation, the interaction energy $\tilde{\cal E}_{\rm int}(\tilde{d})$ between two SQVs with the distance $d = \xi \tilde{d} \gg \xi$ can be estimated as
\begin{eqnarray}
{\cal E}_{\rm int}(d) \simeq - 2 \pi \log(\alpha \tilde{d}),
\label{eq:hydrodynamic-vortex-interaction}
\end{eqnarray}
where we approximate the order parameter as a product of two SQV solutions as 
$
\tilde{\psi}_{\rm int}(\tilde{\boldsymbol{r}},\tilde{d}) \simeq  \tilde{\psi}_1(\tilde{x}-\tilde{d}/2,\tilde{y})\tilde{\psi}_1(\tilde{x}+\tilde{d}/2,\tilde{y})
$.
The dimensionless constant $\alpha = O(1)$ in Eq. \eq{eq:hydrodynamic-vortex-interaction} depends on the lower cutoff of the integration.
The monotonically decreasing structure of $\tilde{\cal E}_{\rm int}(\tilde{d})$ in Eq. \eq{eq:hydrodynamic-vortex-interaction} supports that a DQV is energetically
unfavorable against two SQVs.

We next calculate the interaction energy $\tilde{\cal E}_{\rm int}(\tilde{d})$ more precisely at small $\tilde{d}$.
A naive estimation of $\tilde{\cal E}_{\rm int}$ can be done by the product state
\begin{eqnarray*}
\tilde{\psi}_{\rm int} \simeq \tilde{f}_1(\tilde{x}-\tilde{d}/2,\tilde{y}) \tilde{f}_1(\tilde{x}+\tilde{d}/2,\tilde{y}) e^{i \{ \theta (\tilde{x} - \tilde{d}/2) + \theta(\tilde{x} + \tilde{d}/2) \}},
\end{eqnarray*}
where $\tilde{f}_1(\tilde{x},\tilde{y})$ is the
solution of the amplitude $\tilde{f}(\tilde{\boldsymbol{r}})$ for a SQV at the center $\tilde{x} = \tilde{y} = 0$.
Although the product of $\tilde{f}_1$ for two SQVs should be replaced by $\tilde{f}_2$ for a DQV ,
 we do not consider this change within our naive estimation.
Within the Pad\'e approximation, $\tilde{f}_1(\tilde{x},\tilde{y})$ can be obtained as
\begin{eqnarray}
\tilde{f}_1(\tilde{x},\tilde{y}) \simeq \sqrt{\frac{a_1 \tilde{r}^2 + a_2 \tilde{r}^4}{1 + b_1 \tilde{r}^2 + a_2 \tilde{r}^4}}
\label{eq:Pade-approximation}
\end{eqnarray}
with $\tilde{r}^2=\tilde{x}^2+\tilde{y}^2$.
Here, $a_1$, $a_2$, and $b_1$ satisfy $a_1 = (73 + 3 \sqrt{201}) / 176$, $a_2 = (6 + \sqrt{201}) / 132$, and $b_1 = (21 + \sqrt{201}) / 48$.
For the product state with the amplitude $f_1$ under the Pad\'e approximation, the interaction energy $\tilde{\cal E}_{\rm int}(\tilde{d})$ around $\tilde{d} = 0$ becomes
\begin{eqnarray}
\tilde{\cal E}_{\rm int}(\tilde{d})= - 0.327 \tilde{d}^2 + O(\tilde{d}^4).
\end{eqnarray}
The flat structure of $\tilde{\cal E}_{\rm int}$ at $\tilde{d} = 0$ as $d \tilde{\cal E}_{\rm int} / d \tilde{d} |_{\tilde{d} = 0} = 0$ implies that a DQV is marginally unstable.

\subsubsection*{Comparison with the numerical results}

\begin{figure}[htb]
\centering
\includegraphics[width=1. \linewidth]{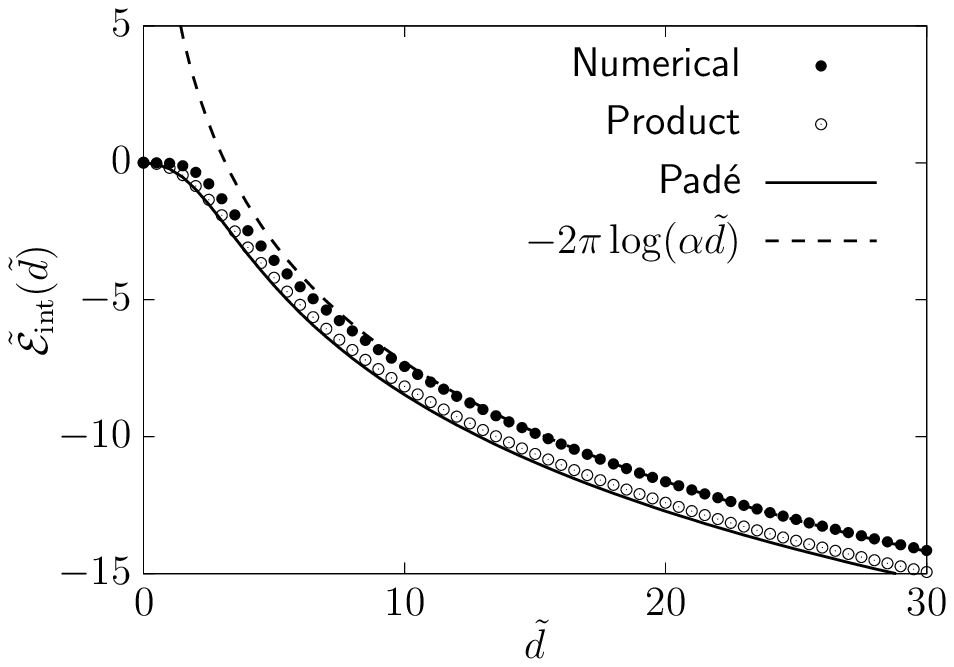}
\caption{
\label{fig:vortex-energy}
Interaction energy $\tilde{\cal E}_{\rm int}(\tilde{d})$ for two SQVs.
Closed circles show the numerically obtained values.
Open circles and the solid line show values given by the product of two SQV solutions obtained numerically and by the Pad\'e-approximation \eq{eq:Pade-approximation} respectively.
Dashed line shows values in Eq. \eq{eq:hydrodynamic-vortex-interaction}, where $\alpha \simeq 0.319$ is chosen as a fitting parameter.
}
\end{figure}

Figure \ref{fig:vortex-energy} shows the interaction energy $\tilde{\cal E}_{\rm int}$ given by the product state with the Pad\'e approximation \eq{eq:Pade-approximation} (solid line),
that with the numerically obtained amplitude $f_1$ (open circles), and Eq. \eq{eq:hydrodynamic-vortex-interaction} (dashed line).
The most precise result (closed circles) are obtained from the ansatz
\begin{eqnarray*}
\tilde{\psi}_{\rm int} = \tilde{f}_{\rm int}(\tilde{\boldsymbol{r}},d) e^{i \{ \theta(\tilde{x} - \tilde{d}/2) + \theta(\tilde{x} + \tilde{d}/2) \}}
\end{eqnarray*}
by numerically computing $\tilde{f}_{\rm int}(\tilde{\boldsymbol{r}},d)$.
The exact interaction energy $\tilde{\cal E}_{\rm int}(\tilde{d})$ at $\tilde{d} = 0$ is flatter than approximated ones with the product states.
All datas converge to the same behavior except for constants at large $\tilde{d}$.

\subsection{Technical description on the numerical analysis}\label{ASec:Tech}

\subsubsection*{Dimensionless equations and the boundary conditions}

The stationary solution of a DQV was obtained by employing the method of steepest descent for the Gross--Pitaevskii (GP) equation.
By rescaling the order parameter amplitude and the radial coordinate as $f=\sqrt{\mu/g}\tilde{f}$ and $r=\xi \tilde{r}$, respectively,
 the GP equation for the DQV state is reduced to
\begin{eqnarray*}
\left[ -\frac{1}{2}\frac{d^2}{d\tilde{r}^2} -\frac{1}{2\tilde{r}}\frac{d}{d\tilde{r}}  - 1+\frac{2}{ \tilde{r}^2}+\tilde{f}^2 \right] \tilde{f} = 0.
\end{eqnarray*}
This equation was solved numerically under
the boundary conditions, $\tilde{f}(\tilde{r}=0)=0$ and $\left. \frac{d\tilde{f}}{d\tilde{r}} \right|_{\tilde{r}=\tilde{R}}=0$.

Similarly, the Bogoliubov--de Gennes (BdG) equation \eq{Eq_radialBdG} can be reduced to a dimensionless form.
The solved equation is
\begin{eqnarray}
\tilde{\omega} \vec{u}= \left[
\begin{array}{cc}
\tilde{h}_+ & -\tilde{f}^2 \\
\tilde{f}^2 & -\tilde{h}_- \\
\end{array} 
\right] \vec{u}
\end{eqnarray}
with $\tilde{\omega}=\frac{\hbar \omega}{\mu}$ and $\tilde{h}_{\pm}=-\frac{1}{2}\frac{d^2}{d\tilde{r}^2} -\frac{1}{2\tilde{r}}\frac{d}{d\tilde{r}}  - 1+\frac{(2\pm m)^2}{2\tilde{r}^2}+2\tilde{f}^2$.
The boundary condition at $\tilde{r}=\tilde{R}$ is $\left. \frac{d\tilde{u}}{d\tilde{r}} \right|_{\tilde{r}=\tilde{R}}=\left. \frac{d\tilde{v}}{d\tilde{r}} \right|_{\tilde{r}=\tilde{R}}=0$.
At $\tilde{r}=0$, we employ $u(\tilde{r}=0)=0$ for $m=-2$ and $v(\tilde{r}=0)=0$ for $m=2$, otherwise $\left. \frac{d\tilde{u}}{d\tilde{r}} \right|_{\tilde{r}=0}=\left. \frac{d\tilde{v}}{d\tilde{r}} \right|_{\tilde{r}=0}=0$.
The rescaled BdG equation was solved numerically by using the Linear Algebra PACKage (LAPACK).
The numerical plots of \Fig{Fig_R_w} were obtained for a mesh size $\Delta r=0.2\xi$ in finite difference methods.

\subsubsection*{Spatial profile around the vortex core}\label{ASec:Vcore}

Because of the symmetry of \Eq{Eq_radialBdG},
if there is a solution (i) ($\omega, m, u, v$) with a complex frequency $\omega$,
we have always other three solutions (ii) ($-\omega, -m, v, u$), (iii) ($\omega^*, m, u^*, v^*$), and (iv) ($-\omega^*, -m, v^*, u^*$).
Consider a  mode (i) with ${\rm Im}(\omega)>0$. Then, 
the mode (iv) with ${\rm Im}(-\omega^*)>0$ is also amplified, while we have ${\rm Im}(-\omega)<0$ for (ii) and ${\rm Im}(\omega^*)<0$ for (iii). 
The solution (iv) is physically identical to the partner (i), because the two solutions yield the same fluctuation $\propto \delta\psi$.
For our problem, we may consider only (i) with ${\rm Im}(\omega)>0$ by neglecting the solutions (ii-iv).

\begin{figure}
\begin{center}
\includegraphics[width=1.0 \linewidth,keepaspectratio]{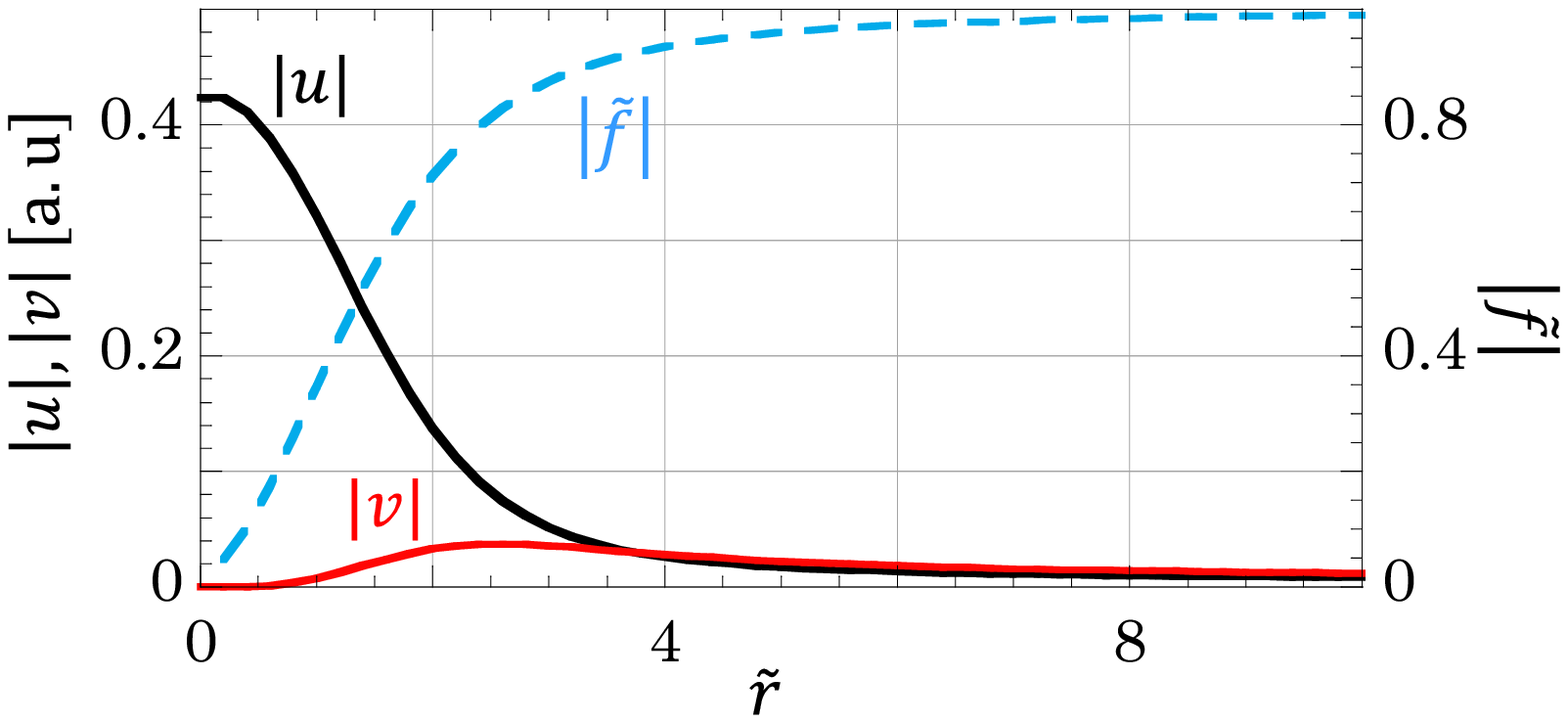}
\end{center}
\vspace{-5mm}
\caption{
The radial profile of the amplitudes $|u|$ and $|v|$ of the instability mode ($m=-2$) for $\tilde{R}=1638.4$.
The dashed curve represents the rescaled profile  $|\tilde{f}(\tilde{r})|$ of the order parameter amplitude.
} 
\label{Fig_r_u_core}
\end{figure}

\Figure{Fig_r_u_core} shows the radial profile of amplitudes $|u|$ and $|v|$ for the instability mode ($m=-2$).
In the vicinity of the vortex core of a DQV ($r/\xi \lesssim 1$),
the asymptotic behaviors of the excitation wave functions are given as $u\propto \(r/\xi\)^{2+m}\to {\rm const.}$ and $v \propto \(r/\xi\)^{2-m}\to 0$ due to their effective centrifugal potentials $\frac{\hbar^2(2+ m)^2}{2m_{\rm a}r^2}\to 0$ and $\frac{\hbar^2(2- m)^2}{2mr^2}=\frac{8\hbar^2}{m_{\rm a}r^2}\to \infty $ in the BdG Hamiltonian $\hat{h}$, respectively.
There, the density fluctuation due to the collective excitation is given by $\delta n \approx \left| \delta\psi \right|^2 \to \left| u \right|^2$ with $f\propto (r/\xi)^2$.

\subsubsection*{The asymptotic values of $\tilde{\omega}_{\rm R}$ and $\tilde{\omega}_{\rm I}$}

\begin{figure}
\begin{center}
\includegraphics[width=1. \linewidth,keepaspectratio]{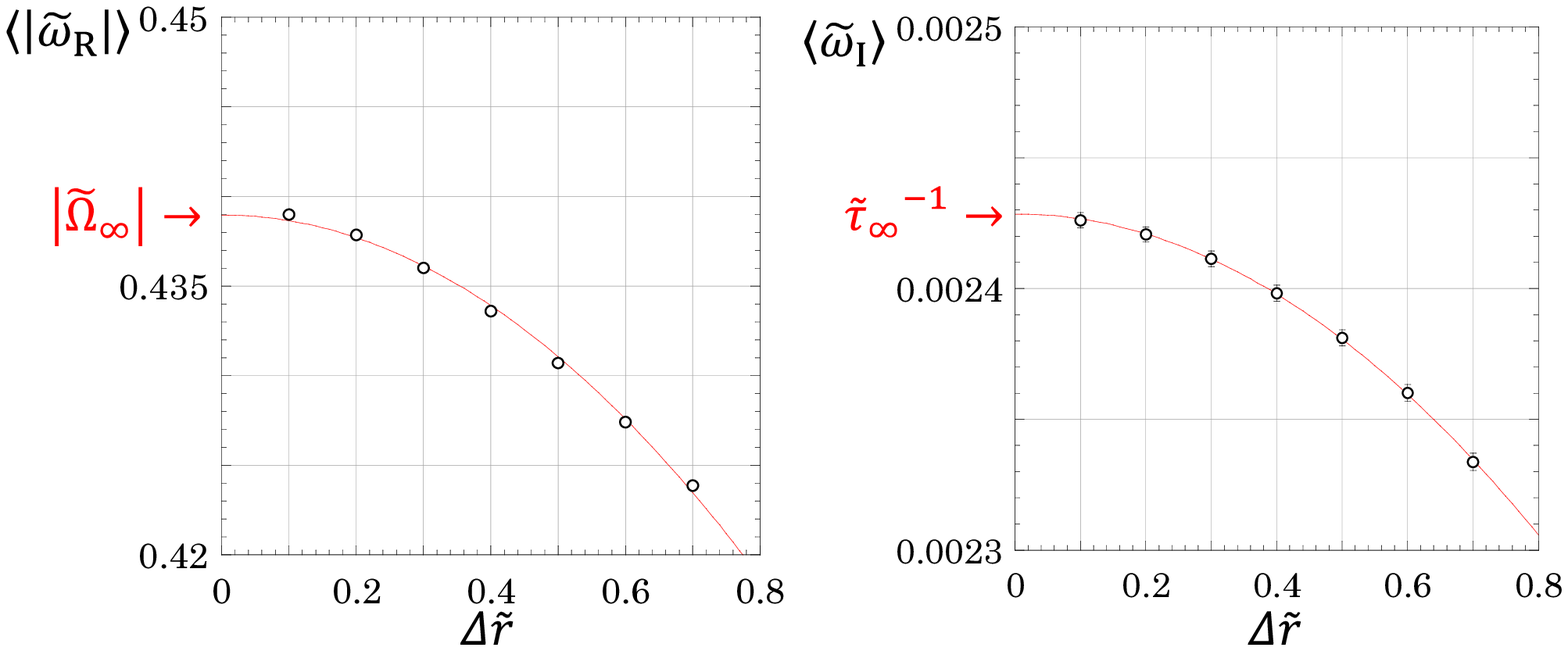}
\end{center}
\vspace{-5mm}
\caption{
The $\Delta \tilde{r}$-dependence of $\<\tilde{\omega}_{\rm R}\>$ and $\<\tilde{\omega}_{\rm I}\>$.
The error bars shows the amplitude $\delta\tilde{\omega}_{\rm R,I}$ of the sinusoidal oscillation within the large-$\tilde{r}$ range ($1630 \leq \tilde{r} \leq 1638$).
The bars are barely seen behind the marks.
} 
\label{Fig_Dr_wRI}
\end{figure}

We have determined the asymptotic values, $\tilde{\Omega}_\infty$ and $\tilde{\tau}_\infty^{-1}$, by considering the dependence of the values $\tilde{\omega}_{\rm R}(<0)$ and $\tilde{\omega}_{\rm I}$ on the numerical grid size $\Delta \tilde{r}$ within the range $1630 \leq \tilde{R} \leq 1638$ (see \Fig{Fig_Dr_wRI}).
When $\tilde{R}$ is sufficiently large,
$\left|\tilde{\omega}_{\rm R}\right|$ ($\tilde{\omega}_{\rm I}$) is simply oscillating like a sinusoidal function of $\tilde{R}$ around its averaged value $\<\left|\tilde{\omega}_{\rm R}\right|\>$ ($\<\tilde{\omega}_{\rm I}\>$) with a small amplitude $\delta\tilde{\omega}_{\rm R,I}$.
The asymptotic value 
$\tilde{\Omega}_\infty=-0.438969 \pm 0.000002$ 
($\tilde{\tau}_\infty^{-1}=0.002429 \pm 0.000002$) 
of $\<\tilde{\omega}_{\rm R}\>$ ($\<\tilde{\omega}_{\rm I}\>$) for the limit $\Delta \tilde{r} \to 0$ is determined by fitting the plot with a quadratic function, $\<\left|\tilde{\omega}_{\rm R}\right|\>=a_{\rm R} \Delta \tilde{r}^2 +\left|\tilde{\Omega}_\infty\right|$ ($\<\tilde{\omega}_{\rm I}\>=a_{\rm I} \Delta \tilde{r}^2 +\tilde{\tau}_\infty^{-1} $) with the method of least squares:
$a_{\rm R}=-0.031606 \pm 0.000007$ and $a_{\rm I}=-0.000192 \pm 0.000007$.
The errors are computed by regarding the small amplitude $\delta\tilde{\omega}_{\rm R,I}$ as the error of the numerical data.

\subsection{The two-mode approximation}\label{ASec:TwoMode}
Let us derive the expression of Eq.~(\ref{omega_per}). The perturbed and unperturbated 
states obey the BdG equations: 
\begin{eqnarray}
\hbar \omega \vec{u} &=& \hat{h}_{\lambda} \vec{u} = ( \hat{h}_{0} + \delta \hat{h} )  \vec{u}, \label{perturba} \\
\hbar \omega_{\alpha} \vec{u}_{\alpha} &=& \hat{h}_{0} \vec{u}_{\alpha},
\end{eqnarray}
respectively. By inserting the expansion $\vec{u} = \sum_{\alpha} {\cal C}_{\alpha} \vec{u}_{\alpha}$ 
into Eq.~(\ref{perturba}), we have
\begin{eqnarray}
\hbar \omega \sum_{\alpha} {\cal C}_{\alpha} \vec{u}_{\alpha} = \sum_{\alpha} \hbar \omega_{\alpha} {\cal C}_{\alpha} \vec{u}_{\alpha} 
+  \sum_{\alpha} {\cal C}_{\alpha} \delta \hat{h} \vec{u}_{\alpha}.
\end{eqnarray}  
By changing the suffix $\alpha$ to $\beta$, multiplying $\vec{u}_{\alpha}^{\dagger} \hat{\sigma}_z$ from the left side, 
and integrating by $2\pi \int_{0}^{\infty} r dr$, we get 
\begin{eqnarray}
\hbar \omega \sum_{\beta} {\cal C}_{\beta} {\cal N}_{\alpha \beta} &=& \sum_{\beta} \hbar \omega_{\beta} {\cal C}_{\beta} {\cal N}_{\alpha \beta} \nonumber \\ 
&+& \sum_{\beta} {\cal C}_{\beta}  \int_{0}^{\infty} 2 \pi r dr \vec{u}_{\alpha}^{\dagger} \hat{\sigma}_z \delta \hat{h} \vec{u}_{\beta}. \label{tptptpoyu}
\end{eqnarray} 
The normalization factor ${\cal N}_{\alpha \beta}$ is written as ${\cal N}_{\alpha \alpha} \delta_{\alpha \beta}$ 
with ${\cal N}_{\alpha \alpha} = \pm 1$ and the Kronecker's delta $\delta_{\alpha \beta}$. Then, dividing Eq.~(\ref{tptptpoyu}) by ${\cal N}_{\alpha \alpha}$, 
we have 
\begin{eqnarray}
\hbar \omega {\cal C}_{\alpha} =\hbar \omega_{\alpha} {\cal C}_{\alpha} + \lambda \sum_{\beta} {\cal C}_{\beta} {\cal W}_{\alpha \beta},  \label{eigenchada}
\end{eqnarray} 
where we defined
\begin{eqnarray}
{\cal W}_{\alpha \beta} = \frac{2\pi}{\lambda {\cal N}_{\alpha \alpha}} \int_{0}^{\infty} r dr \vec{u}_{\alpha}^{\dagger} \hat{\sigma}_z \delta \hat{h} \vec{u}_{\beta}. 
\end{eqnarray}

Here, we introduce the two-mode approximation by taking only $\alpha = 1,2$ to analyze \Eq{eigenchada}. 
The eigenvalue equation is given by 
\begin{eqnarray*}
\hbar \omega \left( 
\begin{array}{c}
{\cal C}_1 \\
{\cal C}_2
\end{array}
\right) = \left( 
\begin{array}{cc}
\hbar \omega_1 + \lambda {\cal W}_{11} & \lambda {\cal W}_{12} \\
 \lambda {\cal W}_{21} & \hbar \omega_2 + \lambda {\cal W}_{22}
 \end{array}
\right) \left(
\begin{array}{c}
{\cal C}_1 \\
{\cal C}_2
\end{array}
\right) .
\end{eqnarray*}
By using the notation 
$\tilde{\varepsilon}_\alpha = (\hbar \omega_{\alpha} + \lambda {\cal W}_{\alpha \alpha}) {\cal N}_{\alpha \alpha} / \mu$ 
and the relation 
$W_{21} = W_{12} ^{\ast}$, the secular equation is written as 
\begin{eqnarray*}
\left({\cal N}_{11} \tilde{\varepsilon}_1 - \tilde{\omega} \right) \left( {\cal N}_{22} \tilde{\varepsilon}_2 - \tilde{\omega} \right) - \lambda^2 \left| \frac{{\cal W}_{12}}{\mu} \right|^2 = 0,
\end{eqnarray*}
whose solution is 
\begin{eqnarray*}
\tilde{\omega} = \frac{{\cal N}_{11} \tilde{\varepsilon}_1 + {\cal N}_{22} \tilde{\varepsilon}_2}{2} \pm 
\sqrt{  \left( \frac{{\cal N}_{11} \tilde{\varepsilon}_1 - {\cal N}_{22} \tilde{\varepsilon}_2}{2} \right)^2 -  \left|\frac{\lambda{\cal W}_{12}}{\mu} \right|^2 }.
\end{eqnarray*}
For ${\cal N}_{11} = -1$ and ${\cal N}_{22} = 1$, this form is consistent with \Eq{omega_per} by introducing 
$\tilde{W}_{\rm mix}^2 = \lambda^2 \left| {\cal W}_{12}/ \mu \right|^2$.

\subsection{The semi-classical approximation}\label{ASec:SemiC}
Starting from the equation $\( E - E_+ \) \( E + E_- \)+g^2f^4=0$ within the zeroth order 
approximation, we can calculate the eigenenergy $E$. Here, $E_\pm$ is given by 
\begin{eqnarray}
E_\pm=\frac{P_r^2}{2m_{\rm a}}+\frac{\(M\pm L\)^2}{2m_{\rm a}r^2}+2gf^2-\mu.
\end{eqnarray}
When we consider the bulk region far from the vortex core, the density profile is 
approximately written as $g f^2 \approx \mu - L^2/(2 m_\mathrm{a} r^2)$, so that 
\begin{eqnarray}
E_\pm \approx \frac{1}{2m_{\rm a}} \left( P_r^2+\frac{M^2}{r^2} \right)+gf^2 \pm \frac{LM}{m_{\rm a}r^2}.
\end{eqnarray}
Furthermore, we neglect the higher order term $O\(r^{-2}\)$, having then $\mu \approx g f^2$ and $E_{\pm} = P_r^2/(2m_\mathrm{a})+ \mu$. 
We eventually get  
\begin{eqnarray}
E^2 = \frac{P_r^2}{2m_\mathrm{a}} \left(  \frac{P_r^2}{2m_\mathrm{a}}  + 2 \mu \right).
\label{usudisp}
\end{eqnarray}

Let us consider the situation in which $E$ is a complex value. 
When $E$ is real, the momentum $P_r$ is also a real value, according to \Eq{usudisp}. 
When $E$ is complex, the momentum $P_r$ should be 
written as $P_r = \hbar k + i \hbar \kappa$. 
When $E_{\rm R}\equiv {\rm Re}(E) \gg E_{\rm I}\equiv {\rm Im}(E)$,
it is reasonable to assume as $k \gg \kappa$. 
Substituting $E=E_{\rm R}+iE_{\rm I}$ and $P_r = \hbar k + i \hbar \kappa$ into \Eq{usudisp}, 
and comparing the real and imaginary parts of the both sides of the equation, we have 
\begin{eqnarray}
E_{\rm R}^2 - E_{\rm I}^2 &=& \frac{\hbar^2 (k^2 -\kappa^2)}{2 m_\mathrm{a}}  \left[  \frac{\hbar^2 (k^2 -\kappa^2)}{2 m_\mathrm{a}}  + 2 \mu \right] - \frac{\hbar^4k^2 \kappa^2}{m_\mathrm{a}^2} , \nonumber \\
2 E_{\rm R} E_{\rm I} &=& \frac{\hbar^2 k \kappa}{m_\mathrm{a}} \left[ \frac{\hbar^2(k^2 - \kappa^2)}{m_\mathrm{a}}  + 2 \mu \right]. \nonumber 
\end{eqnarray}
By introducing the dimensionless values $\tilde{\omega}_{\rm R,I} = E_{\rm R,I}/\mu$, $\tilde{k} = k \xi$, $\tilde{\kappa} = \kappa \xi$ with 
$\xi = \hbar/\sqrt{m_\mathrm{a} \mu}$, the above equation can be written as 
\begin{eqnarray}
\tilde{\omega}_{\rm R}^2 - \tilde{\omega}_{\rm I}^2 &=& \frac{\tilde{k}^2 - \tilde{\kappa}^2}{2}  \left[ \frac{\tilde{k}^2 - \tilde{\kappa}^2}{2}  + 2 \right] - \tilde{k}^2 \tilde{\kappa}^2, \nonumber \\ 
 2 \tilde{\omega}_{\rm R} \tilde{\omega}_{\rm I} &=& \tilde{k} \tilde{\kappa} (\tilde{k}^2 - \tilde{\kappa}^2 + 2). \nonumber 
\end{eqnarray}
For $E_{\rm R}^2 \gg E_{\rm I}^2$ and $k^2 \gg \kappa^2$, the above equations can be further reduced to  
\begin{eqnarray}
\tilde{\omega}_{\rm R}^2 &\approx& \frac{\tilde{k}^2}{2}   \left( \frac{\tilde{k}^2}{2}   + 2 \right), 
 \nonumber \\ 
 2 \tilde{\omega}_{\rm R} \tilde{\omega}_{\rm I} & \approx & \tilde{k} \tilde{\kappa} (\tilde{k}^2  + 2). \nonumber 
\end{eqnarray}
By solving these equations, we can get 
\begin{eqnarray}
\tilde{k} &\approx& \pm |\tilde{\omega}_{\rm R}| \left( 1- \frac{\tilde{\omega}_{\rm R}^2}{8}\right), \\
\tilde{\kappa} &\approx& \tilde{\omega}_{\rm I} \left( 1- \frac{3\tilde{\omega}_{\rm R}^2}{8}\right).
\end{eqnarray}




\begin{thebibliography}{99}
\bibitem{BubbleWidth} The width of the $j$th bubble is defined as $2\delta \tilde{R}_j=2| \tilde{R}-\tilde{R}_j|$ such that $\tilde{\omega}_{\rm I}(\tilde{R})=0$ within the range $\tilde{R}_{j-1}<\tilde{R}<\tilde{R}_{j+1}$.
Assuming $\tilde{\varepsilon}_{\rm ph}(\tilde{k})\approx \tilde{k}_j$ and $\delta \tilde{R}_j \ll \tilde{R}_j$, which are well-satisfied in our case,
 we have $\delta \tilde{R}_j=2\tilde{R}_j|\tilde{W}_{\rm mix}/ \tilde{\varepsilon}_{\rm core}|$.
 Then, the real part within the bubble is written as $|\tilde{\omega}_{\rm R}|\approx |\tilde{\varepsilon}_{\rm core}|(3/2-\tilde{R}/\tilde{R}_j)$.
The neighboring bubbles are overlapped if $\tilde{R}_{j+1}-\tilde{R}_j < \delta\tilde{R}_{j+1}+\delta\tilde{R}_j$,
 which is satisfied for $\tilde{R}_j \sim |\tilde{W}_{\rm mix}|^{-1}$, namely $10^2 \lesssim \tilde{R}_j \lesssim 10^3$.
These behaviors are consistent with the plots in \Fig{Fig_R_w}.
\bibitem{PowerLaw} Although the power law for $\tilde{R} \lesssim 500$ must reflect the finite-size-effect through the mixing interaction $\tilde{W}_{\rm mix}$, we have never succeed a quantitative evaluation of $\tilde{W}_{\rm mix}$ analytically.
\end{thebibliography}
\end{document}